\documentclass[aps,onecolumn,superscriptaddress]{revtex4}
\usepackage{epsfig}
\usepackage{amssymb,amsmath,amsfonts,amsthm,graphicx}
\graphicspath{{./Figures/}}

\newcommand{\beq}{\begin{equation}}
\newcommand{\eeq}{\end{equation}}
\newcommand{\bea}{\begin{array}}
\newcommand{\eea}{\end{array}}
\newcommand{\beqa}{\begin{eqnarray}}
\newcommand{\eeqa}{\end{eqnarray}}

\newcommand{\ch}{\cosh}
\newcommand{\sh}{\sinh}

\def\beqa{\begin{eqnarray}}
\def\eeqa{\end{eqnarray}}

\begin{document}
\title{Study of QCD Phase Diagram with Non-Zero Chiral Chemical Potential}

\author{V.~V.~ Braguta}
\affiliation{Institute for High Energy Physics NRC "Kurchatov Institute", Protvino, 142281 Russian Federation}
\affiliation{Institute of Theoretical and Experimental Physics, 117259 Moscow, Russia}
\affiliation{Far Eastern Federal University,  School of Natural Sciences, 690950 Vladivostok, Russia} 
\affiliation{Moscow Institute of Physics and Technology, Dolgoprudny, 141700 Russia}

\author{E.-M.~Ilgenfritz}
\affiliation{Joint Institute for Nuclear Research, BLTP, 141980 Dubna, Russia}

\author{A.~Yu.~Kotov}
\affiliation{Institute of Theoretical and Experimental Physics, 117259 Moscow, Russia}
\affiliation{National Research Nuclear University MEPhI (Moscow Engineering Physics Institute), Moscow 115409, Russia}

\author{B.~Petersson}
\affiliation{Humboldt-Universit\"at zu Berlin, Institut f\"ur Physik,
12489 Berlin, Germany}

\author{S.~A.~Skinderev}
\affiliation{Institute of Theoretical and Experimental Physics, 117259 Moscow, Russia}

\begin{abstract}
In this paper we report on lattice simulations of $SU(3)$--QCD with 
non-zero chiral chemical potential.
We focus on the influence of the chiral chemical potential on the 
confinement/deconfinement phase transition 
and the breaking/restoration of chiral symmetry. 
The simulation is carried out with dynamical Wilson fermions.  
We find that the critical temperature rises as the 
chiral chemical potential grows.
\end{abstract}

\maketitle

\section{Introduction}

The aim of relativistic heavy ion collisions is to study the phase 
structure of QCD in a more or less unknown region and to explore exotic 
properties of the phases becoming accessible at high densities of energy 
and baryonic number.
Understanding the interplay of strong electromagnetic fields with the 
basic mechanisms of QCD has turned out as similarly interesting direction of 
research since 2008~\cite{Kharzeev:2007jp}, although not primarily targeted 
as the high matter density itself, which is characterized by the baryonic 
chemical potential $\mu_B$. This interplay, in particular in the quark-gluon 
phase, makes its study even more interesting. Here the chiral chemical 
potential $\mu_5$ is of central importance.

Indeed, in non-central collisions the strongest magnetic fields on earth 
($qB \sim {\cal O}(1 .. 15) m_\pi^2$) are created~\cite{Skokov:2009qp} 
at RHIC and LHC with a lifetime of few ${\mathrm{fm}}/c$ in each collision. 
There exist also electric field fluctuations of a similar order of 
magnitude~\cite{Bzdak:2011yy}. If electric and magnetic fields are coupled 
through the Ohmic and anomalous conductivities of the plasma, the magnetic 
field due to the spectators is practically unaltered~\cite{McLerran:2013hla}.

The coupling between color and electromagnetic fields belongs to field of
phenomena possible due to the axial 
anomaly~\cite{Adler:1969gk,Bell:1969ts,Bardeen:1974ry}.    
Besides the well-known role it is playing in the QCD vacuum at $T=0$, 
it has a constitutive role under the extremal conditions which exist in 
the intermediate non-hadronic phase as well as in certain condensed matter
systems~\cite{Kharzeev:2015kna} as recently has been noticed. 

The violation of $U_A(1)$ symmetry
(which is not spontaneously broken but rather due to the topological vacuum 
structure at $T=0$) is well-known in the form of mass splittings between 
hadrons~\cite{Witten:1979vv,Veneziano:1979ec} (and signalled by further 
observables 
like $\Delta = \langle \pi \pi \rangle - \langle \delta \delta \rangle$, the
splitting between the $\pi$ and $\delta$ norms, the $T$ dependence of the 
Dirac spectral density, and of the topological susceptibility $\chi_{\rm top}$).
The $U_A(1)$ symmetry will be restored in the temperature range up to few 
times $T_c$.
How this happens in detail in real QCD in contrast to pure Yang-Mills 
theory~\cite{Bornyakov:2013iva}, is under intensive study on the lattice
\cite{Buchoff:2013nra,Sharma:2013nva,Aoki:2012yj,Cossu:2013uua,Brandt:2012sk,
Brandt:2013mba}.

Instead of topological charge appearing (at the infrared scale) in the form 
of Euclidean instantons or dyons~\cite{Ilgenfritz:2013oda}), the axial anomaly 
will play a dynamical role during the short time of  existence of liberated 
quarks. 
Mesoscopic domains of (almost classical) strongly CP violating gluon fields 
are created during the initial stadium of a collision (with color electric 
and magnetic fields ${\vec E^a} || {\vec B^a} || {\rm collision~axis}$). 
This is believed to lead to the accumulation of a chiral imbalance between 
quarks, $n_5=n(\mathrm{lefthanded})-n(\mathrm{righthanded}) \ne 0$
(with random sign and strength). 
The analogon of instanton tunneling at high temperature, the sphaleron 
transitions over-the-barrier (frequent at $T > T_c$) give rise to a 
Chern-Simons diffusion rate $\Gamma \propto T^4$) which is insufficient to 
completely wash out the chiral imbalance existing during the lifetime of the 
high-temperature phase. 
This, together with the magnetic field ${\vec B}$ mentioned before, gives 
rise to anomalous transport of electric current along the magnetic field  
(the so-called Chiral Magnetic Effect, for a recent report 
see Ref.~\cite{Kharzeev:2015znc}), as a remnant of the temporal and local 
violation of P and CP invariance). 
Thus, a charge asymmetry w.r.t. the collision plane remains as effect of 
the electric current ${\vec J} \propto \mu_5~{\vec H}$ flowing while the 
deconfined phase has existed. This is the crucial role of $\mu_5$. One can 
also note other phenomena in media with the chiral 
imbalance~\cite{Sadofyev:2015hxa,Sadofyev:2015hxa}.

Of course, a precise scenario has to be the result of a space-time transport
simulation. In an intermediate step, lattice QCD should explore the 
multidimensional phase diagram, including (beyond temperature) one or two 
chemical potentials (baryonic $\mu_B$ and/or $\mu_5$) and the magnetic field
strength. In a first generation of lattice studies, the effect of an external 
magnetic field on the temperature of the chiral and/or deconfining transition 
has been studied~\cite{D'Elia:2010nq,Bali:2011qj,Ilgenfritz:2012fw} 
in isolation. This has led to the discovery of the interplay of opposite 
tendencies (magnetic catalysis and inverse catalysis of the chiral condensate) 
at different temperatures. The phase diagram of QCD in a magnetic field has 
been reviewed in Ref.~\cite{Andersen:2014xxa}. 

Baryonic density $n_B = (n({\mathrm{quarks}})-n({\mathrm{antiquarks}}))/3$, 
modelled by $\mu_B \ne 0$, still remains difficult to simulate (if not 
$\mu_B << T$) except for $SU(2)$. For this model system a simulation with 
$N_f=2$ flavors of staggered fermions has been recently 
performed~\cite{Braguta:2015lzp,Braguta:2015cta}, following earlier work for 
$SU(2)$ (with Wilson fermions) which has been summarized in 
Ref.~\cite{Cotter:2012mb}. Chiral imbalance, however, 
$n_5=n(\mathrm{lefthanded})-n(\mathrm{righthanded}) \ne 0$ (expected in the 
result of the chiral anomaly), can be modelled by a chiral chemical potential 
$\mu_5$ without a sign problem. It would be very interesting to study the 
simultaneous influence of magnetic field and chiral imbalance on the 
transition temperature. In such studies, as soon as a magnetic field 
${\vec H}$ is included, the induced electric current 
${\vec J} \propto \mu_5 {\vec B}$ would be among the most interesting 
observables as proposed in Ref.~\cite{Yamamoto:2012bi}.

In a few recent papers~\cite{Braguta:2015sqa,Braguta:2015zta}, our 
collaboration has begun to study the effect of non-vanishing chiral 
chemical potential $\mu_5$ on the transition temperature. We did this first  
for two-color Yang-Mills theory coupled to staggered fermions without rooting
(four flavors). With the present paper we report our results obtained within 
an extension of our codes to $SU(3)$ Yang-Mills theory coupled to two flavors 
of Wilson fermions. Preliminary results have been already reported 
at ``Lattice 2015''.

Section 2 is devoted to the details of the calculation. 
In section 3 we present our results, and in section 4 we discuss them. 
In an appendix we study renormalization properties of the chiral condensate 
with nonzero chiral chemical potential.

\section{Details of the simulation.}

In the simulation we used the $SU(3)$ one-plaquette gauge action and a Wilson 
fermionic action with two degenerate quark flavors. The Dirac operator with 
non-zero chiral chemical potential has the form~\cite{Yamamoto:2011gk}  	
\begin{equation}\begin{split}
	D_{xy}=1-\kappa\sum\limits_i\left((1-\gamma_i)U_{i}(x)\delta_{x+i,y}+(1+\gamma_i)U_{i}^{\dag}(y)\delta_{x-i,y}\right)-\\
	-\kappa\left((1-\gamma_4e^{\mu_5a\gamma_5})U_{4}(x)\delta_{x+4,y}+(1+\gamma_4e^{-\mu_5a\gamma_5})U_{4}^{\dag}(y)\delta_{x-4,y}\right)
\label{eq:fermions}
\end{split}\end{equation}

Here $\mu_5$ enters through an additional exponential factor for timelike link 
variables. In the naive continuum limit $a \to 0$, the fermion action with the 
Dirac operator (\ref{eq:fermions}) corresponds to the fermion action with a 
chiral chemical potential:
\begin{equation}\begin{split}
	S_f^{(cont)}=\int d^4x \bar{\psi}(\partial_{\mu}\gamma_{\mu}+igA_{\mu}\gamma_{\mu}+m+\mu_5\gamma_5\gamma_0)\psi
\end{split}\end{equation}
The Wilson Dirac operator (\ref{eq:fermions}) is $\gamma_5$-hermitean:
\begin{equation}
	\gamma_5D^{\dag}(\mu_5)\gamma_5=D(\mu_5)
\end{equation}
This property implies that its determinant $\det D(\mu_5)$ is real, and in 
the case of $N_f=2$ fermion flavors it is positive. 
Thus the system has no sign problem and can be simulated by standard Hybrid 
Monte-Carlo methods.

In our simulations we used Wilson fermions since they allow the introduction 
of the chiral chemical potential $\mu_5$ in a local exponential form, what is 
not possible for staggered fermions. Moreover, calculations with dynamical 
Wilson fermions require not so many computational resources.

We have performed simulations on a lattice $4\times16^3$, with $\kappa=0.1665$ 
kept fixed throughout the simulations (performed as $\beta$-scan). 
For $\beta=5.32144$ (in the transition region) the chosen $\kappa$ corresponds 
to a lattice step size $a=0.13$~fm and to a  pion mass $m_{\pi}=418$~MeV. 
Our measured observables are the Polyakov loop, the chiral condensate 
and their respective susceptibilities. The Polyakov loop and its susceptibility
are sensitive to the confinement/deconfinement transition, 
while the chiral condensate and its susceptibility are sensitive to the chiral 
symmetry breaking/restoration aspects.
We present the observables as functions of $\beta$ for our scans at four 
different values of $\mu_5a=0.0,0.25,0.5,0.75$. 

\section{The results of the simulation}

\begin{figure*}[h!]
\begin{tabular}{cc}
\includegraphics[scale=0.65,clip=false]{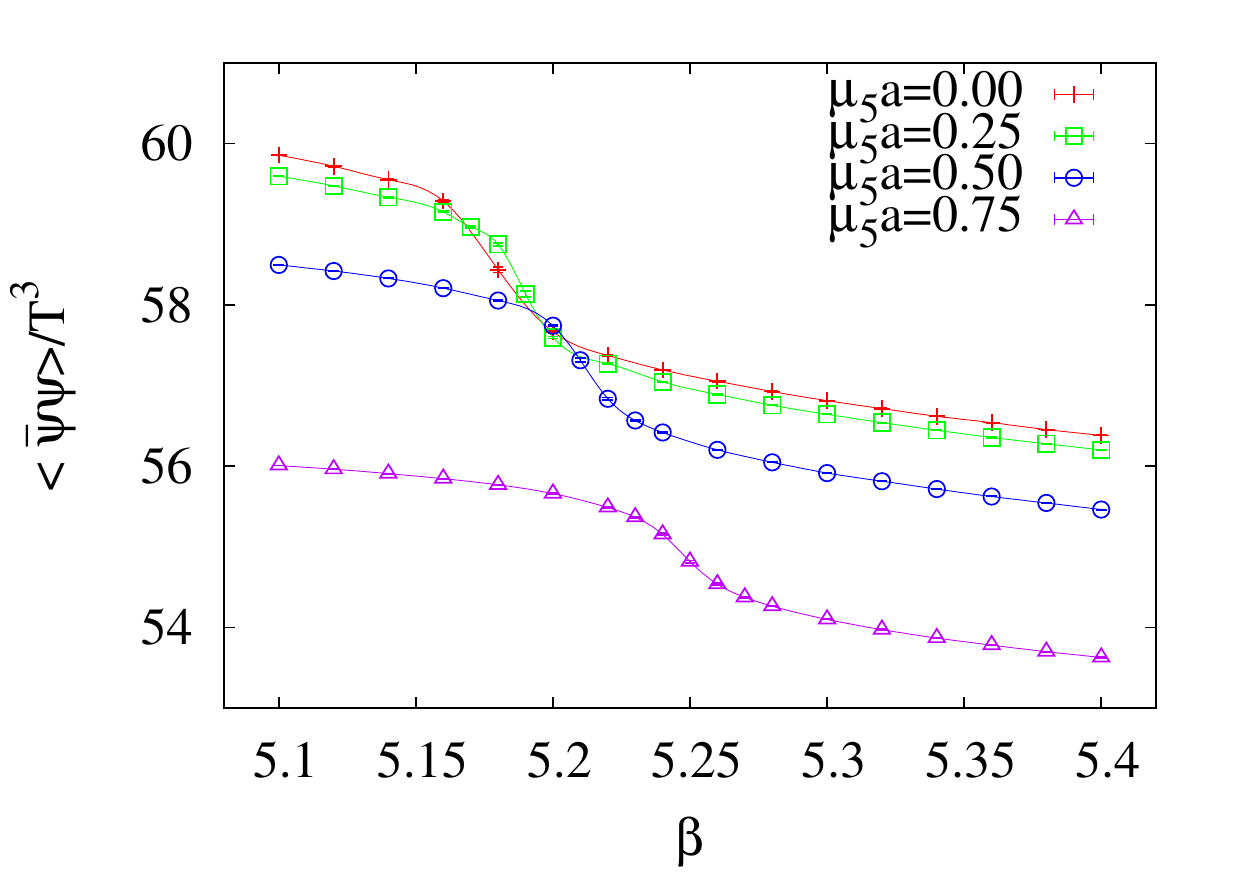} &
\includegraphics[scale=0.65,clip=false]{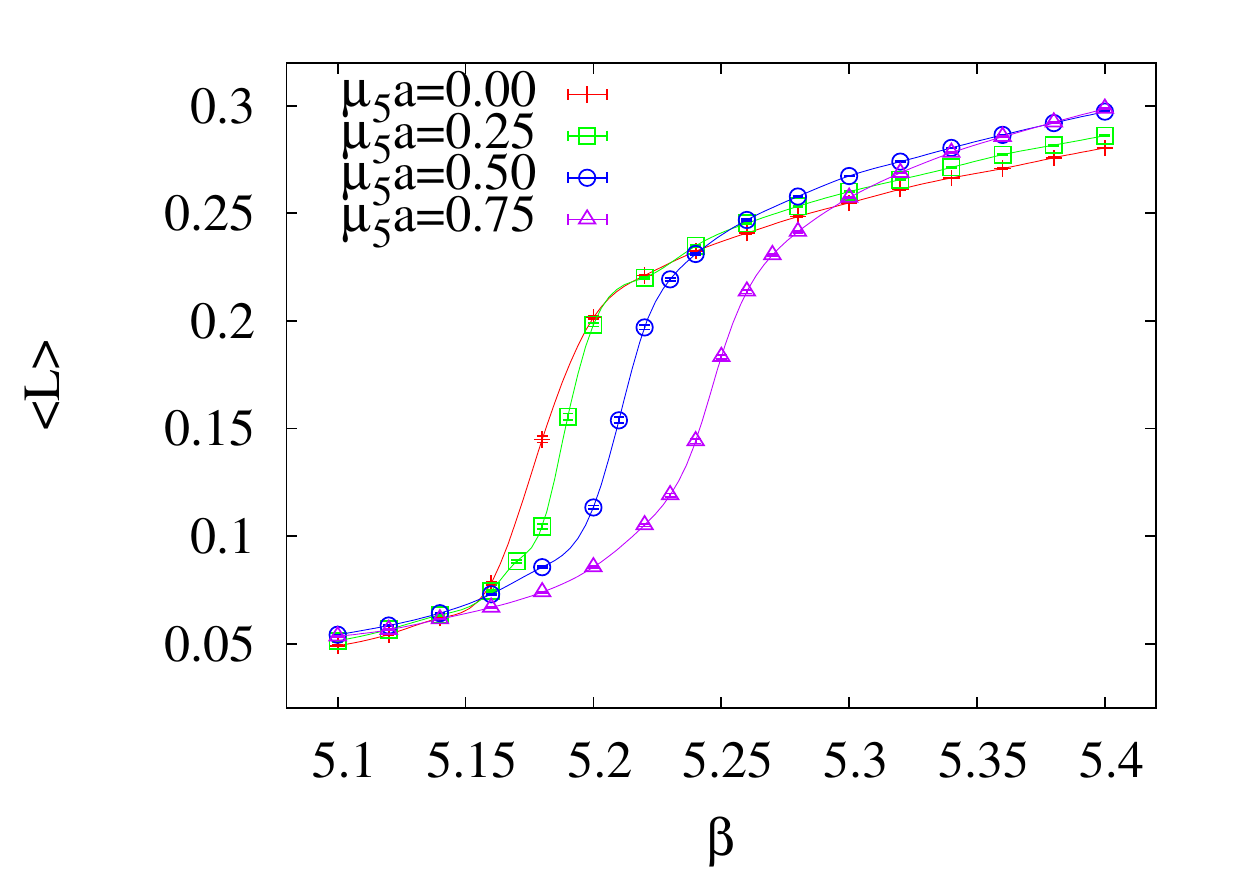} \\
chiral condensate & Polyakov loop
\end{tabular}
\caption{Chiral condensate and Polyakov loop versus 
$\beta$ for four values of $\mu_5a$. The lattice size is $4 \times 16^3$, 
the hopping parameter is $\kappa=0.1665$. Errors are smaller than the data 
point symbols. The curves are to guide the eye.
}
\label{fig:obs}
\end{figure*}

In Figure \ref{fig:obs} we show the Polyakov loop and the chiral condensate 
as functions of $\beta$ for various $\mu_5a$. 
The sharp change of observables indicates the position of the phase transition.
One clearly sees that nonzero chiral chemical potential shifts both phase 
transitions to larger values of $\beta$. This means that the transition 
temperatures increases with $\mu_5$. A splitting between both phase transitions
is not observed.

\begin{figure*}[h!]
\begin{tabular}{cc}
\includegraphics[scale=0.65,clip=false]{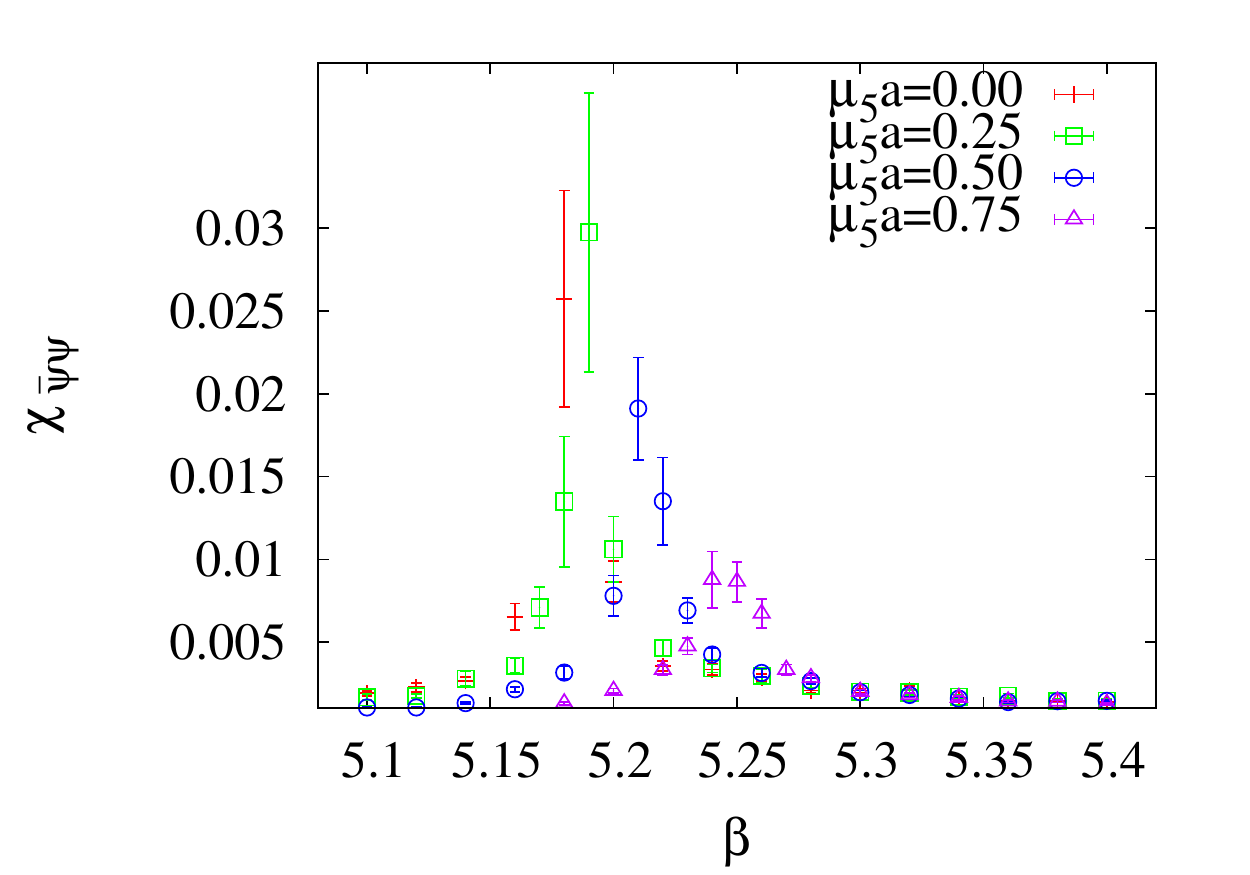} &
\includegraphics[scale=0.65,clip=false]{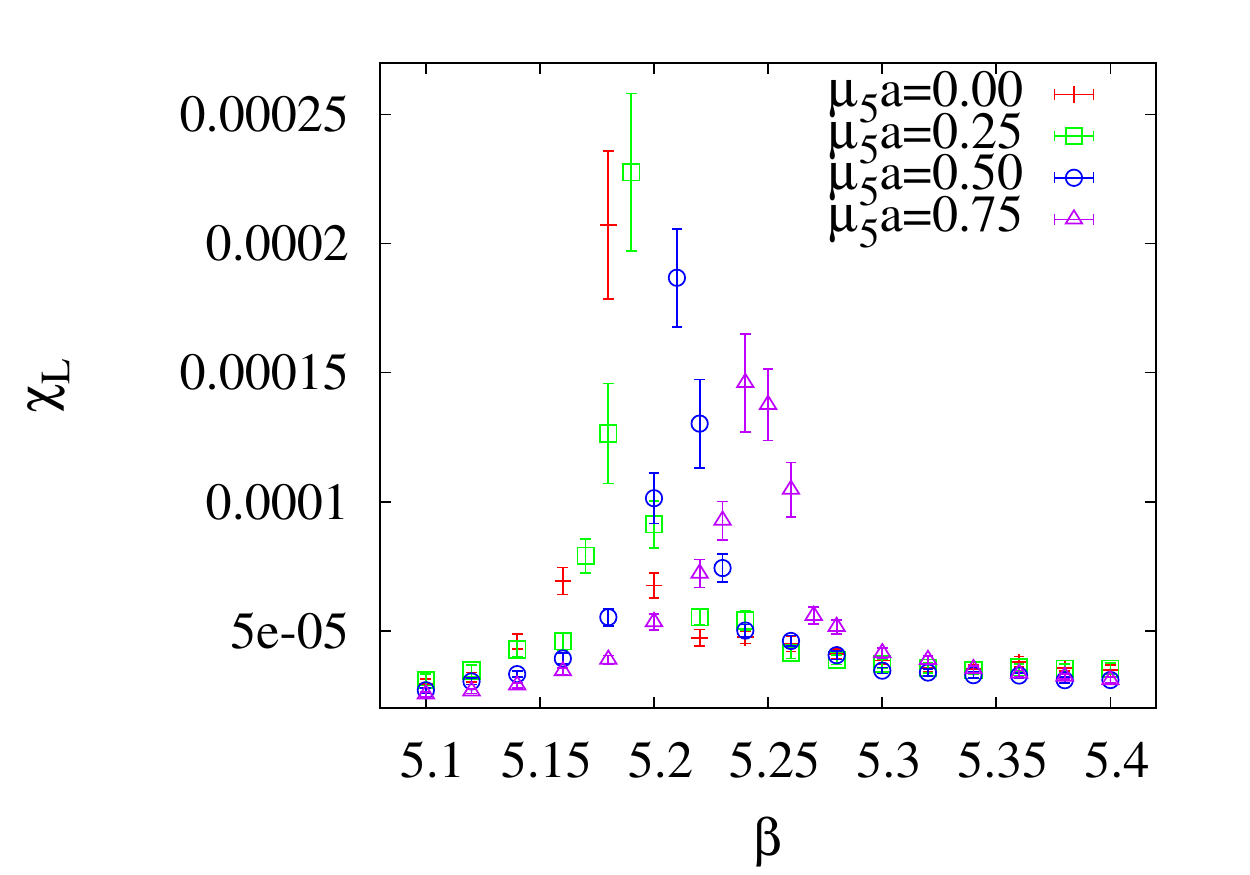} \\
chiral condensate & Polyakov loop
\end{tabular}
\caption{Chiral susceptibility and Polyakov loop susceptibility versus
$\beta$ for four values of $\mu_5a$. The lattice size is $4 \times 16^3$, 
the hopping parameter is $\kappa=0.1665$. 
}
\label{fig:sus}
\end{figure*}

In order to confirm our observation and to quantify our results we also have
measured the chiral susceptibility and the Polyakov loop susceptibility as 
functions of $\beta$. The results are presented in Figure \ref{fig:sus}. 
The peak positions of the susceptibilities correspond to the positions of 
the respective transition. From these observables one can definitely conclude
that the critical temperature grows with increasing $\mu_5$. 

\begin{figure*}[h!]
\includegraphics[scale=0.65,clip=false]{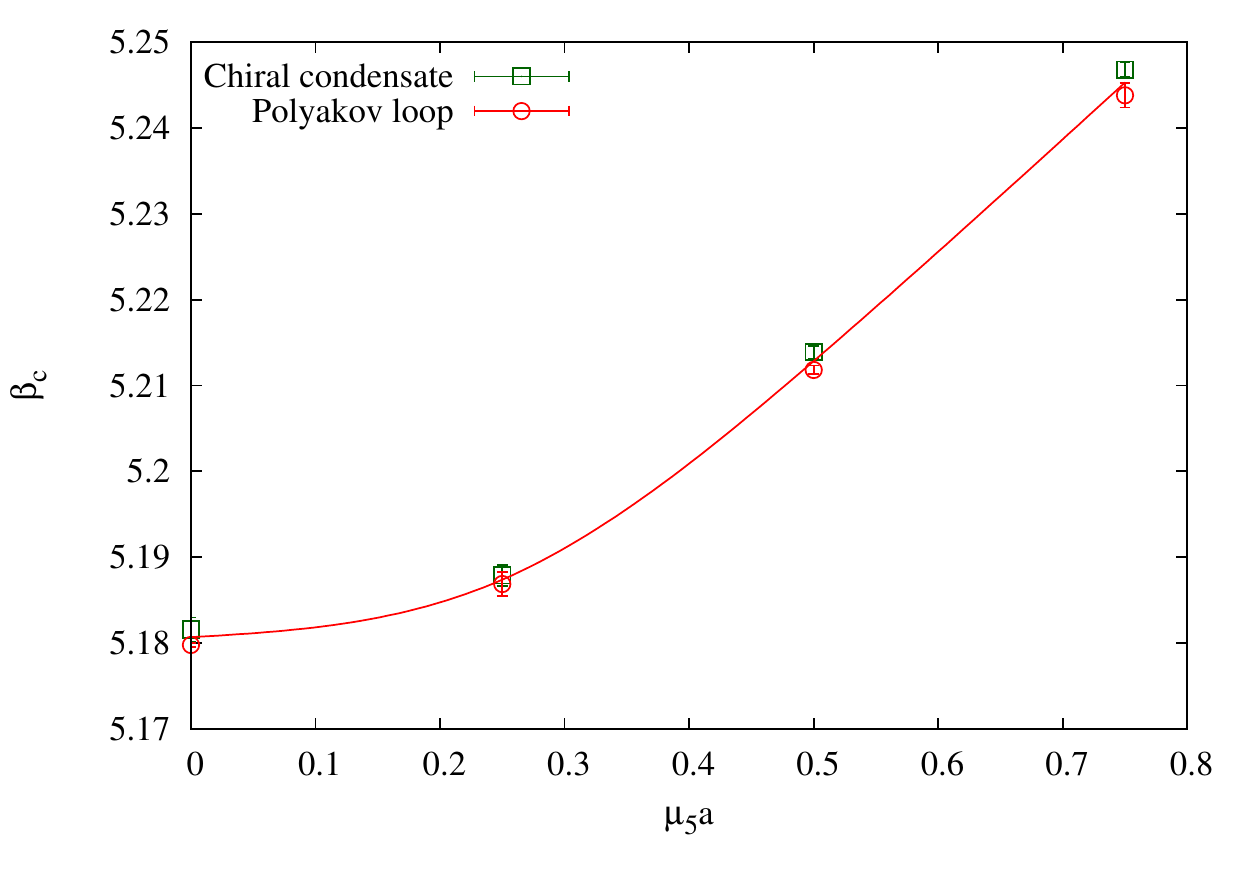} 
\caption{Critical $\beta_c$ for the chiral condensate and the Polyakov loop 
versus $\mu_5a$. The lattice size is $4 \times 16^3$, the hopping parameter
$\kappa=0.1665$. The curve is to guide the eyes.
}
\label{fig:fit}
\end{figure*}

\begin{figure*}[h!]
\includegraphics[scale=0.65,clip=false]{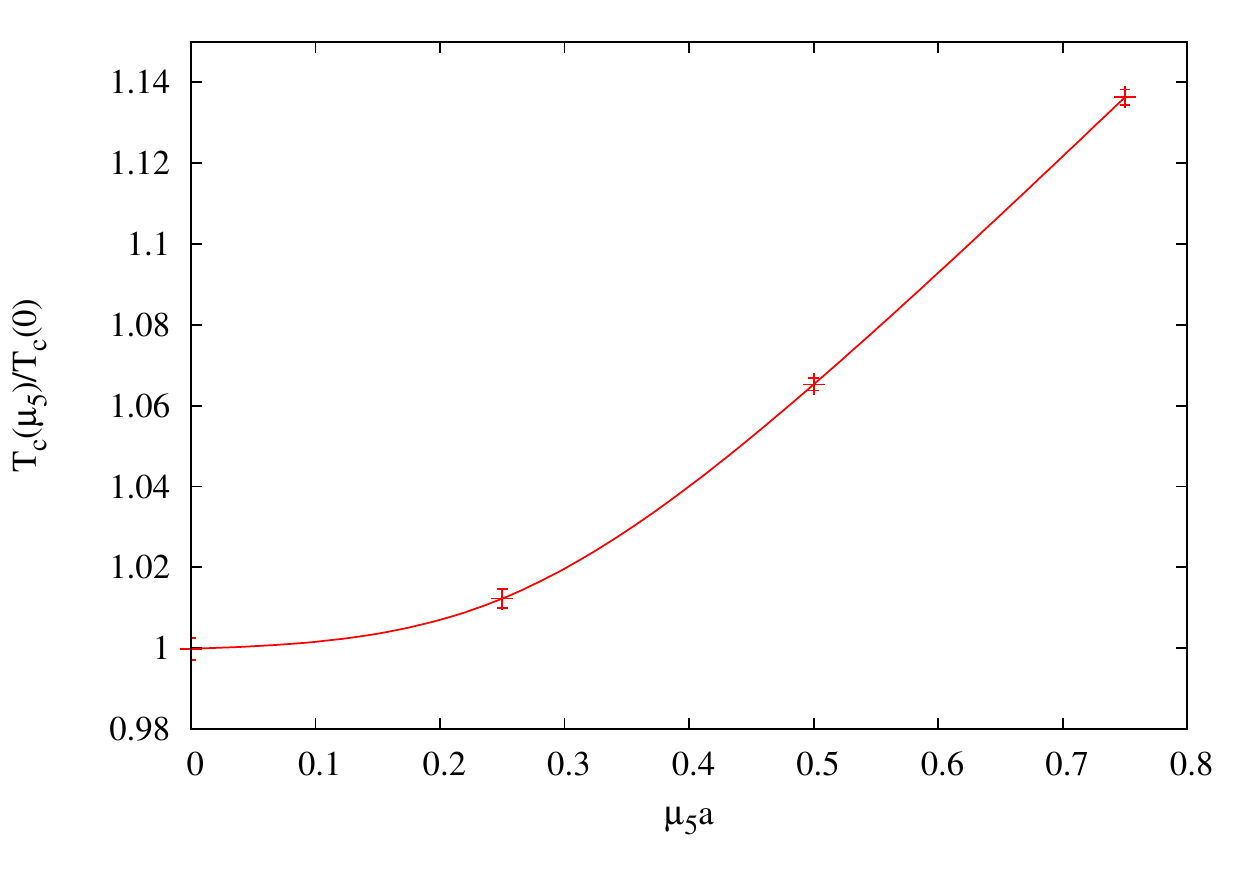} 
\caption{Critical temperature $T_c(\mu_5)/T_c(0)$ for the chiral condensate 
	versus $\mu_5a$. The lattice size is $4 \times 16^3$, the hopping
parameter is $\kappa=0.1665$. The curve is to guide the eyes.
} \label{fig:phase}
\end{figure*}

In order to determine the critical temperatures we fitted the plots for the 
susceptibilities with a Gaussian function: 
$f(\beta)=a_0+a_1e^{-(\beta-\beta_c)^2/\sigma^2}$ (we used 6-7 points near 
the peak). The dependence of the critical $\beta_c$ extracted from the fit 
is shown in Fig. \ref{fig:fit}. For both susceptibilities it can be described 
by a quadratic fit:
\begin{equation}\begin{split}
	\beta_c = 5.18 + 0.12 (\mu_5a)^{2}
\end{split}\end{equation}
Using the two-loop $\beta$-function, we extract the dependence of the critical 
temperature on $\mu_5a$. The resulting phase diagram is presented in 
Fig. \ref{fig:phase}.

At the end of this section it is important to discuss the ultraviolet 
divergences of the chiral condensate and of the Polyakov loop and how they 
affect our results. 
In the paper \cite{Braguta:2015zta} it was shown that the introduction of a 
non-zero chiral chemical potential for staggered fermions leads to an 
additional logarithmic divergence in the chiral condensate 
$\sim \mu_5^2 \log (a)$, whereas it does not lead to an additional 
divergence in the Polyakov loop. In the present work we perform simulations 
with Wilson fermions, and it is important to study the ultraviolet divergences 
which appear in the observables in the case of Wilson fermions. First one can 
repeat all steps in the derivation presented in \cite{Braguta:2015zta} and 
show that there are no additional divergences due to $\mu_5 \neq 0$ for 
Wilson fermions in the Polyakov loop.

In the Appendix we present an analytical calculation of additional divergences 
in the chiral condensate for free Wilson fermions. The results of this study 
allow us to state that there are two additional divegences due to 
$\mu_5 \neq 0$: the first one is logarithmic, $\sim \mu_5^2 \log (a)$ 
and the second one is linear $\sim \mu_5^2/a$.
The logarithmic divergence also exists in the case of staggered 
fermions~\cite{Braguta:2015zta} with the same coefficient. 
The linear divergence is new as compared to staggered fermions. We believe 
that the linear divergence in the chiral condensate appears due to the explicit
chiral symmetry breaking of Wilson fermions. Note that the coefficient in front 
of the linear divergence is negative. So, an increase of the chiral chemical 
potential leads to a decrease of the chiral condensate. It seems that this 
behaviour persists in the interacting case. 
From Fig. \ref{fig:obs} one sees that the curves with larger $\mu_5$ are 
shifted down compared to the curves with smaller $\mu_5$.

Despite the fact that the additional divergences due to $\mu_5 \neq 0$ give 
a considerable contribution to the value of the chiral condensate, we believe 
that there is no influence of the divergences to the position of the 
breaking/restoration chiral symmetry transition for the following reasons. 
First, as was noted above there are no divergences in the Polyakov loop. 
So, if additional divergences had influenced the position of the chiral 
symmetry breaking/restoration transition, it would be visible as a splitting 
between the chiral symmetry breaking/restoration and the 
confinement/deconfinement transitions.
But we don't see such splitting. Notice also that the position of the phase 
transition manifests itself as a peak of the susceptibility. Evidently, there 
is no peak due to the additional ultraviolet divergence. All this allows us to 
state that the conclusions obtained in this paper are not affected by the 
additional ultraviolet $\mu_5$ divergence in the chiral condensate.

\section{Conclusions}

In this paper we have studied the phase diagram of $SU(3)$--QCD with a chiral 
chemical potential within lattice simulations using $N_f=2$ flavors of 
dynamical Wilson fermions. We have calculated the chiral condensate, the 
Polyakov loop, the Polyakov loop susceptibility and the chiral susceptibility 
for different values of the temperature $T$ and the chiral chemical potential 
$\mu_5$ on lattices of size $4 \times 16^3$. 
The main result is that at non-zero values of the chiral chemical potential 
the critical temperatures of the confinement/deconfinement phase transition 
and of the chiral-symmetry breaking/restoration phase transition still 
coincide, and that the common transition is shifted to larger temperatures 
as $\mu_5$ increases. 

In the present paper we have reported the first investigation of the phase 
diagram of $SU(3)$--QCD with a non-vanishing chiral chemical potential. 
In the following, one should extrapolate to physical quark masses and to the 
continuum limit in the hope to confirm the results. This exploration requires 
much larger computational resources and remains as a task for future studies.

Our result is in agrement with a lattice study of the phase diagram of 
$SU(2)$--QCD~\cite{Braguta:2015zta,Braguta:2015sqa}. Notice that the simulation 
described in papers \cite{Braguta:2015zta,Braguta:2015sqa} was carried out with staggered fermions which in the continuum corresponds to $N_f=4$ flavors.
A chiral chemical potential was introduced to the lattice action as an additive
term. So, simulations with different fermion discretization 
and different numbers of colors and flavors give similar results. 
We believe that this is an argument in favour of the statement that 
the critical temperature of QCD--like systems rises as the chiral chemical 
potential is increased. 

The studies of the QCD phase diagram with a chiral chemical potential within 
different effective models have given controversial results. For instance, 
the authors of papers 
\cite{Fukushima:2010fe,Chernodub:2011fr,Gatto:2011wc,Chao:2013qpa,Yu:2014sla} 
have predictied that the transition shifts to smaller temperatures as $\mu_5$ 
increases. At the same time, some of the results obtained in 
papers \cite{Andrianov:2013dta,Yu:2015hym}
imply that the transition shifts to larger temperatures as $\mu_5$ increases.
In paper \cite{Yu:2015hym} it was shown that the results of effective models 
crucially depend on finer details of the models. 

In paper \cite{Braguta:2015unpub} the influence of the chiral chemical 
potential on the chiral symmetry breaking/restoration transition was studied. 
It was shown that the chiral chemical potential enhances the chiral symmetry 
breaking 
and shifts the critical temperature to larger values. The result of our paper
confirms this statement.

Besides effective models, the phase diagram of QCD in the 
$(\mu_5, T)$--plane was studied in papers \cite{Wang:2015tia,Xu:2015vna}
in the framework of Dyson-Schwinger equations. We would like also to mention 
the paper \cite{Hanada:2011jb}. In this paper the authors address the question 
of universality of phase diagrams in QCD and QCD--like theories through the 
large--$N_c$ equivalence. 
The authors of these papers found that the critical temperature rises with 
$\mu_5$. These results agree with ours. 

\section*{Acknowledgements}

The simulations were performed at the ITEP supercomputer. 
The work was supported by the Far Eastern Federal University, 
the Dynasty foundation, by RFBR grants 14-02-01185-a, 15-02-07596-a, 
15-32-21117, and by a grant of the FAIR-Russia Research Center.

\appendix

\section{Ultraviolet divergences in the chiral condensate}

The fermion propagator including the chiral chemical potential for Wilson 
fermions can be written in the following form
\begin{eqnarray}
\nonumber
S^{\alpha\beta}(x,y)&=&\frac{\d elta^{\alpha\beta}}{L_t L_s^3}\sum_{\{k\}}\sum_{s=\pm1}e^{ip(x-y)}\frac{m+\frac{\hat k^2}{2}-i\sum_{i=1}^3\gamma_i \sin k_i-i\gamma_4 \sin k_4 \ch\mu_5-\cos k_4\sh\mu_5\gamma_4\gamma_5}
{\left(m+\frac{\hat k^2}{2}\right)^2+\sin^2 k_4 \ch^2 \mu_5+\left(|{\bf \bar k}|+s\cos k_4\sh\mu_5\right)^2}\times P(s),\\
P(s)&=&\frac{1}{2}\left(1-is\sum_{i=1}^3\frac{\gamma_i \sin k_i}{|{\bf \bar k}|}\gamma_4\gamma_5\right),\nonumber\\
|{\bf \bar k}|&=&\sqrt{\sin^2 k_1+\sin^2 k_2+\sin^2 k_3},\nonumber\\
\hat k_\mu &=& 2 \sin(k_\mu/2),\nonumber\\
k_i &=& \frac{2\pi}{L_s}n_i , i = 1, 2, 3, n_i = 0, ..., L_s - 1,\nonumber\\
k_4 &=& \frac{2\pi}{L_t}n_4 + \frac{\pi}{L_t}, n_4 = 0, ..., L_t - 1.\nonumber
\end{eqnarray}
Here $m$ and $\mu_5$ are mass and chiral chemical potential in lattice units, 
$\alpha$, $\beta$ are color indices, the sum is taken over all possible values 
of ($n_1$, $n_2$, $n_3$, $n_4$) (lattice coordinates).

In the limit $L_s, L_t \to \infty$ the condensate can be written as
\begin{equation}
\label{condensate}
\langle\bar\psi\psi\rangle=Tr[S(x,x)]=6\int_{-\pi}^{\pi}\frac{d^4k}{(2\pi)^4}\sum_{s=\pm1}\frac{m+\frac{\hat k^2}{2}}{(m+\frac{\hat k^2}{2})^2+\sin^2 k_4 \ch^2 \mu_5+(|{\bf \bar k}|+s\cos k_4\sh\mu_5)^2} ;\ . 
\end{equation}
To calculate the integral in formula (\ref{condensate}) we use 
the algebraic method, explained in~\cite{Burgio:1996ji,Capitani:2002mp}.
The Taylor expansion in $\mu_5$ tends to
\beq
\label{integral}
\frac{1}{6}\langle\bar\psi\psi\rangle=2\int_{-\pi}^{\pi}\frac{d^4k}{(2\pi)^4}\frac{m+\frac{r}{2}\hat k^2}{(m+\frac{r}{2}\hat k^2)^2+\sum_\mu \bar k_\mu^2}+
\mu_5^2\left(-\frac{2(m+\frac{\hat k^2}{2})}{\left((m+\frac{\hat k^2}{2})^2+\sum_\mu \bar k_\mu^2\right)^2}+\frac{8\cos^2k_4|{\bf \bar k}|^2(m+\frac{\hat k^2}{2})}{\left((m+\frac{\hat k^2}{2})^2+\sum_\mu \bar k_\mu^2\right)^3}
\right)+O(\mu_5^4). \nonumber
\eeq
The first term in this expression is just the loop integral for 
$\langle\bar\psi\psi\rangle$ without chiral chemical potential. 
The expression for this integral in the limit $a \to 0$ can be written as 
follows
\beqa
\label{term1}
2\int_{-\pi}^{\pi}\frac{d^4k}{(2\pi)^4}\frac{m+\frac{r}{2}\hat k^2}{(m+\frac{r}{2}\hat k^2)^2+\sum_\mu \bar k_\mu^2} &=& c_0+c_1 m+c_2 m^2+m^3\left(\frac{\log m^2}{8\pi^2}+c_3\right)+O(m^4) \\ \nonumber
&& c_0=0.469363,~ c_1=-0,067967,~ c_2=-0,023613,~ c_3= -0,075829. \nonumber
\eeqa
The second term in (\ref{integral}) we can calculate similarly.
The result of the calculation can be written as follows
\beqa
\label{term2}
\int_{-\pi}^{\pi}\frac{d^4k}{(2\pi)^4}\left(-\frac{2(m+\frac{\hat k^2}{2})}{\left((m+\frac{\hat k^2}{2})^2+\sum_\mu \bar k_\mu^2\right)^2}+\frac{8\cos^2k_4|{\bf \bar k}|^2(m+\frac{\hat k^2}{2})}{\left((m+\frac{\hat k^2}{2})^2+\sum_\mu \bar k_\mu^2\right)^3}\right)
&=&c_4+m\left(-\frac{\log m^2}{4\pi^2}+c_5\right)+O(m^2) \\ \nonumber
&& c_4 = -0.010738,~ c_5=0,045999. \nonumber
\eeqa
We have checked the two last formulas numerically. Restoring the lattice 
spacing $a$ in our results we get 
\beq
\langle\bar\psi\psi\rangle=\frac{6c_0}{a^3}+\frac{6c_1 m}{a^2}+\frac{6c_2 m^2}{a}+m^3\left(\frac{3\log(m a)}{2\pi^2}+6c_3\right)+ \\
+ \frac{6c_4 \mu_5^2}{a}+\mu_5^2m\left(-\frac{3\log(m a)}{\pi^2}+6c_5\right).
\label{res_d}
\eeq

The first line in formula (\ref{res_d}) represents the chiral condensate 
without chiral chemical potential, whereas the second line is the contribution 
due to non-zero $\mu_5$. It is seen that there is an additional logarithmic 
divergence due to $\mu_5 \neq 0$. The same divergence with the same coefficient
appears in the case of staggered fermions \cite{Braguta:2015zta}. However, 
there is also a linear divergence $\sim \mu_5^2/a$ which is absent in the case 
of staggered fermions. We believe that this linear divergence in the chiral 
condensate appears due to the explicit chiral symmetry breaking of Wilson 
fermions. Note that the coefficient in front of the linear divergence is 
negative. 
So, an increase of the chiral chemical potential leads to a decrease of the 
chiral condensate what is seen in Fig.\ref{fig:obs}.

\bibliographystyle{unsrt}

\end{document}